\documentclass[12pt]{article}

\usepackage{amsfonts,mathtools,setspace}
\usepackage[textheight=9in,textwidth=6.5in,letterpaper]{geometry}

\numberwithin{equation}{section}

\newcommand{\av}[1]{\left\langle#1\right\rangle}
\newcommand{\br}[1]{\left[#1\right]}
\newcommand{\cu}[1]{\left\{#1\right\}}
\newcommand{\pa}[1]{\left(#1\right)}
\newcommand{\rn}[1]{\av{#1}_R}
\newcommand{\bra}[1]{\left\langle#1\right|}
\newcommand{\ket}[1]{\left|#1\right\rangle}

\newcommand{\e}{\epsilon}
\newcommand{\pd}{\,\partial}
\newcommand{\so}{\mathsf{SO}}
\renewcommand{\O}{\Omega}

\makeatletter

\renewcommand\section{\@startsection {section}{1}{\z@}%
  {-3.5ex \@plus -1ex \@minus -.2ex}%
  {2.3ex \@plus.2ex}%
  {\normalfont\large\bfseries}}

\renewcommand\subsection{\@startsection{subsection}{2}{\z@}%
  {-3.25ex\@plus -1ex \@minus -.2ex}%
  {1.5ex \@plus .2ex}%
  {\normalfont\bfseries}}

\begin{document}

\begin{titlepage}
  \begin{center}

    {\LARGE{\textsc{Quasinormal Quantization in deSitter Spacetime}}}

    \vspace{2cm}

    Daniel L. Jafferis$^\dagger$,
    Alexandru Lupsasca$^\dagger$,
    Vyacheslav Lysov$^\dagger$,
    Gim Seng Ng$^\dagger$\\
    and Andrew Strominger$^{\dagger\ddagger}$

    \footnote[0]{$\dagger$
      Center for the Fundamental Laws of Nature,
      Harvard University,
      Cambridge, MA, USA}

    \footnote[0]{$\ddagger$
      Radcliffe Institute for Advanced Study,
      Cambridge, MA, USA}

    \vspace*{2cm}

    \begin{abstract}
      A scalar field in four-dimensional deSitter spacetime (dS$_4$)
      has quasinormal modes which are singular on the past horizon of
      the south pole and decay exponentially towards the future. These
      are found to lie in two complex highest-weight representations
      of the dS$_4$ isometry group $\so(4,1)$. The Klein-Gordon norm
      cannot be used for quantization of these modes because it
      diverges. However a modified `R-norm', which involves reflection
      across the equator of a spatial $S^3$ slice, is nonsingular. The
      quasinormal modes are shown to provide a complete orthogonal
      basis with respect to the R-norm. Adopting the associated
      R-adjoint effectively transforms $\so(4,1)$ to the symmetry
      group $\so(3,2)$ of a 2+1-dimensional CFT. It is further shown
      that the conventional Euclidean vacuum may be defined as the
      state annihilated by half of the quasinormal modes, and the
      Euclidean Green function obtained from a simple mode
      sum. Quasinormal quantization contrasts with some conventional
      approaches in that it maintains manifest dS-invariance
      throughout. The results are expected to generalize to other
      dimensions and spins.
    \end{abstract}

  \end{center}
\end{titlepage}

\setcounter{page}{1}

\tableofcontents

\vspace{1cm}

\section{Introduction}

In this paper we present a new and potentially useful approach to an
old problem: the quantization of a scalar field in four-dimensional de
Sitter spacetime (dS$_4$), which has an $\so(4,1)$ isometry group.
One standard approach begins with the spherical harmonics of the $S^3$
spatial sections, and proceeds by solving the wave equation for the
time-dependent modes. Linear combinations of these modes that are
nonsingular under a certain analytic continuation are then identified
as the Euclidean modes and used to define the quantum Euclidean
vacuum. The vacuum so constructed exhibits manifest
$\so(4)$-invariance and can also be shown to possess the full
$\so(4,1)$ symmetry of dS$_4$. Another common approach singles out the
southern causal diamond and relies on a special Killing vector field,
denoted $L_0$, which generates southern Killing time and whose
corresponding eigenmodes have real frequency $\omega$. This
construction displays manifest $\so(3)\times\so(1,1)$ symmetry and
again leads to the dS-invariant Euclidean vacuum. The modes employed
in these and similar constructions are {\it not} in $\so(4,1)$
multiplets and hence $\so(4,1)$-invariance of the final expressions is
nontrivial. For example, the action of the dS$_4$ isometries on the
southern diamond $L_0$ eigenmodes shifts the frequency by {\it
  imaginary} integer multiples of $2\pi/\ell$ (where $\ell$ is the dS
radius) while the usual southern diamond modes all have real
frequencies.

It is natural to adopt scalar modes which lie in highest-weight
representations of $\so(4,1)$ and therefore boast manifest
dS-invariance. These turn out to be nothing but the quasinormal modes
of the southern diamond, which have complex $L_0$ eigenvalues and
comprise four real or two complex highest-weight
representations.\footnote{Interesting work on the normalizability and
  completeness of quasinormal modes for black holes can be found in
  \cite{Berti:2009kk,Ching:1995rt,Beyer:1998nu,Nollert:1998ys}.} They
are singular on the past horizon and decay exponentially towards the
future, as opposed to the conventional southern diamond modes which
oscillate everywhere. In order to quantize in a quasinormal mode
basis, a norm is needed. The singularities on the past horizon render
the Klein-Gordon norm singular, which is presumably why the
quasinormal modes have not typically been used for
quantization. However a variety of other equally suitable norms {\it
  have} been employed for various reasons in dS \cite{Sanchez:1986gr,
  Gibbons:1986dd,Witten:2001kn,Balasubramanian:2002zh,Bousso:2001mw,
  Parikh:2002py,Ng:2012xp,Anninos:2012qw}. One of them -- the
so-called R-norm \cite{Ng:2012xp} -- differs from the Klein-Gordon
norm by the insertion of a spatial reflection through the equator of
the $S^3$ slice, thereby exchanging the north and south poles. We
demonstrate that the R-norm is finite for quasinormal modes and hence
suitable for quantization. We also show that the Euclidean vacuum has
the simple and manifestly dS-invariant definition as the state
annihilated by two of the four sets of quasinormal modes. Moreover the
Euclidean Green function is shown, as anticipated in
\cite{Anninos:2011af}, to be obtainable from a simple sum over the
quasinormal modes. We caution the reader that quasinormal modes have
singularities on the past horizon which we regulate with an
$i\e$-prescription. Our statements about completeness and mode sums
depend on taking the $\e\to0$ limit at the end of our calculations.

The real Killing vectors which generate the dS isometries have an
$\so(4,1)$ Lie bracket algebra and are antihermitian with respect to
the Klein-Gordon norm. However they have mixed hermiticity under the
R-norm. Multiplication by appropriate factors of $i$ produces complex
Killing vector fields which are antihermitian under the R-norm. The
Lie algebra of these R-antihermitian vector fields turns out to
generate $\so(3,2)$, which is precisely the symmetry group of a
2+1-dimensional CFT, and the transformed notion of hermiticity is
exactly the one conventionally employed when studying CFT$_3$ on the
Euclidean plane \cite{Bousso:2001mw}. Hence this $\so(4,1)\to\so(3,2)$
transformation, and the use of quasinormal modes, fits naturally
within the dS$_4$/CFT$_3$ correspondence
\cite{Witten:2001kn,Hull:1998vg,Strominger:2001pn,Strominger:2001gp,
  Maldacena:2002vr, Harlow:2011ke,Anninos:2011ui}.\footnote{For every
  bulk scalar $\Phi$ one expects a dual operator ${\cal O}$ in the
  boundary CFT$_3$. The bulk state with one quantum in the lowest
  quasinormal mode is dual to the CFT$_3$ state associated to an
  ${\cal O}$ insertion at the north pole of the $S^3$ at
  $\mathcal{I}^+$, and the descendants fill out $\so(3,2)$
  representations on both sides of the duality \cite{Ng:2012xp}.}

This paper is organized as follows. In section 2, we begin by
reviewing the standard global $S^3$ modes and the construction of
Euclidean modes and Green functions. In section 3, we show that the
quasinormal modes comprise the highest-weight modes and their
descendants, specializing for simplicity to the case of conformal mass
$m^2\ell^2=2$. Then in section 4, the modified R-norm and its
properties are presented. Next, in section 5 we prove that half the
quasinormal modes are Euclidean modes and demonstrate their
completeness by deriving the Euclidean Green function from a
quasinormal mode sum. In section 6 we generalize these results to the
case of light scalars with $m^2\ell^2\le9/4$. Finally, in section 7 we
isolate quasinormal modes that vanish in the northern or southern
diamonds -- the analogues of Rindler modes in Minkowski space. These
might eventually be useful for understanding the thermal nature of
physics in a single dS causal diamond, but we do not pursue this
direction further herein.

In addition, in Appendix \ref{appendix:KV} we provide the explicit
forms of dS$_4$ Killing vectors as well as their commutation
relations. This is followed by Appendix \ref{appendix:Norm}, which
computes the norm of spherically symmetric descendants using the
$\so(4,1)$ algebra, and Appendix \ref{appendix:Green}, which provides
details on the Euclidean two-point function evaluated on the south
pole observer's worldline.

We expect our discussion to generalize to the case of heavy scalars
with $m^2\ell^2>9/4$ as well as other dimensions and spin.

\section{$\so(4)$-invariant global mode decomposition}

In this section we describe the standard dS$_4$ mode decomposition in
terms of the spherical harmonics of the $S^3$ spatial sections. These
modes are regular everywhere on dS$_4$ and sometimes referred to as
`global modes'.

We will work in the dS$_4$ global coordinates $x=(t,\psi,\theta,\phi)$
with line element
\begin{equation}
  \frac{ds^2}{\ell^2}
  =-dt^2+\cosh^2{t}\,d\O_3^2
  =-dt^2+\cosh^2{t}\br{d\psi^2
    +\sin^2{\psi}\pa{d\theta^2+\sin^2{\theta}\,d\phi^2}},
\end{equation}
where $\O=(\psi,\theta,\phi)$ are coordinates on the global $S^3$
slices. We denote the north and south pole by
\begin{equation}
  \O_{SP}\sim\psi=0,\qquad\O_{NP}\sim\psi=\pi.
\end{equation}
In this coordinate system, the dS-invariant distance function
$P(x;x')$ is given by
\begin{equation}
  P(t,\O;t',\O')
  =\cosh{t}\cosh{t'}\cos\Theta_3(\O,\O')-\sinh{t}\sinh{t'},
\end{equation}
where $\Theta_3(\O,\O')$ denotes the geodesic distance function on
$S^3$ and
\begin{equation}
  \cos\Theta_3(\O,\O')
  =\cos\psi\cos\psi'+\sin\psi\sin\psi'
  \br{\cos\theta\cos\theta'+\sin\theta\sin\theta'\cos(\phi-\phi')}.
\end{equation}
Following the notation of \cite{Ng:2012xp}, solutions of the wave
equation
\begin{equation}
  \pa{\nabla^2-m^2}\Phi=0
\end{equation}
may be expanded in representations of the $\so(4)$ rotations of the
$S^3$ spatial slice at fixed $t$:
\begin{equation}
  \Phi_{Lj}(x)=y_L(t)Y_{Lj}(\O).
\end{equation}
These have total $\so(4)$ angular momentum $L$ and spin labeled by the
multi-index $j$. The $S^3$ spherical harmonics $Y_{Lj}$ obey the
identities
\begin{eqnarray}
  Y^*_{Lj}(\O)&=&(-)^LY_{Lj}(\O)\ =\ Y_{Lj}(\O_A),\cr
  D^2Y_{Lj}(\O)&=&-L(L+2)Y_{Lj}(\O),\cr
  \int_{S^3}\!d^3\O\sqrt{h}\ Y^*_{Lj}(\O)Y_{L'j'}(\O)
  &=&\delta_{L,L'}\delta_{j,j'},\cr
  \sum Y^*_{Lj}(\O)Y_{Lj}(\O')&=&\frac{1}{\sqrt{h}}\delta^3(\O-\O'),
\end{eqnarray}
where $\O_A$ denotes the antipodal point of $\O$, while
$\sqrt{h}=\sin^2{\psi}\sin{\theta}$ and $D^2$ are the measure and
Laplacian on the unit $S^3$, respectively. Here and hereafter, $\sum$
denotes summation over all allowed values of $L$ and the multi-index
$j$. The time dependence $y_L(t)$ is then governed by the differential
equation
\begin{equation}
  \pd_t^2y_L+3\tanh{t}\pd_ty_L+\br{m^2\ell^2+\frac{L(L+2)}{\cosh^2t}}y_L
  =0.
\end{equation}
The general solution has the $\mathcal{I}^+$ falloff
\begin{equation}
  y_L\to e^{-h_\pm t},\qquad
  h_\pm=\frac{3}{2}\pm\sqrt{\frac{9}{4}-m^2\ell^2}.
\end{equation}
For the time being, we restrict our attention to the case
$m^2\ell^2=2$, which corresponds to a conformally coupled scalar with
\begin{equation}
  h_+=2,\qquad h_-=1.
\end{equation}
The case of generic mass is qualitatively similar but with algebraic
functions replaced by hypergeometric ones. We give the correspondingly
more involved formulae in section 6.

The so-called Euclidean modes, which define the vacuum, are those
which remain nonsingular on the southern hemisphere when dS$_4$ is
analytically continued to $S^4$. In other words, they are defined by
the condition
\begin{equation}
  y^E_L\pa{t=-{\frac{i\pi}{2}}}=\mathrm{nonsingular}.
\end{equation}
Explicitly, these modes are \cite{Ng:2012xp}:
\begin{equation}
  y^E_L=\frac{2^{L+1}}{\sqrt{2L+2}}
  \frac{\cosh^L{t}\,e^{-(L+1)t}}{(1-ie^{-t})^{2L+2}}.
\end{equation}
Note that they are singular on the northern hemisphere at
$t=i\pi/2$. In terms of the Klein-Gordon inner product on global $S^3$
slices,
\begin{equation}
  \av{\Phi_1,\Phi_2}_{KG}
  \equiv i\int_{S^3}\!d^3\Sigma^\mu\
  \Phi^{*}_{1}\overleftrightarrow{\pd_\mu}\Phi_{2},
\end{equation}
we have normalized the modes such that
\begin{equation}
  \av{\Phi_{Lj}^E,\Phi_{L'j'}^E}_{KG}=\delta_{LL',jj'}.
\end{equation}
Using these modes, one can define the Euclidean vacuum by the
condition
\begin{equation}
  \av{\Phi_{Lj}^E,\hat\Phi}_{KG}\ket{0_E}=0,
\end{equation}
where $\hat\Phi$ is the quantum field operator. Since the modes
$\Phi_{Lj}^E$ are not $\so(4,1)$-invariant, it is not immediately
obvious that the Euclidean vacuum is dS-invariant, but this can be
checked explicitly. The Wightman function is
\begin{equation}
  G_{E}(x;x')
  \equiv\bra{0_E}\hat\Phi(x)\hat\Phi(x')\ket{0_E}
  =\sum\Phi_{Lj}^E(x)\Phi_{Lj}^{E*}(x').
\end{equation}
Using the $i\e$-prescription, this may be expressed in terms of the
dS-invariant distance function $P(x;x')$ as
\begin{equation}
  \label{eq:GE}
  G_{E}(x;x')=\frac{1}{8\pi^2}\frac{1}{1-P(x;x')+is(x;x')\e},
\end{equation}
where $s(x;x')>0$ if $x$ lies in the future of $x'$ and $s(x;x')<0$
otherwise.

If we rewrite $P(x;x')$ in terms of the coordinates $X$ on the
embedding 5D manifold with Minkowski spacetime metric $\eta$ (in which
dS$_4$ is just the hyperboloid $\eta_{\mu\nu}X^\mu X^\nu=\ell^2$),
then we can represent $s(x;x')$ by \cite{Spradlin:2001pw}
\begin{equation}
  \label{eq:ss}
  s(X;Y)\equiv X^0-Y^0.
\end{equation}
Note that this is exactly the same as sending $X^0-Y^0\rightarrow
X^0-Y^0-i\e$, since this latter choice of $i\e$-prescription shifts
$P(X;Y)=\eta_{\mu\nu}X^\mu Y^\nu/\ell^2$ as follows:
\begin{equation}
  P(X;Y)\rightarrow P(X;Y)-i\e(X^0-Y^0).
\end{equation}
dS$_4$ has 10 real Killing vectors which, letting $k\in\cu{1,2,3}$, we
will refer to as the dilation $L_0$, the 3 boosts $M_{k}-M_{-k}$ and
the 6 $\so(4)$ rotation generators $J_k$ and $M_k+M_{-k}$. Their
explicit forms are given in Appendix \ref{appendix:KV}. The global
modes indexed by $L$ transform in the $(L,L)$ representation of
$\so(4)$ with quadratic Casimir $L(L+2)$, but they are not in definite
$\so(4,1)$ representations. In particular, acting arbitrarily many
times with the $L_0$ raising or lowering operators $M_{\pm k}$ gives a
nonzero result, so they are in representations with unbounded
$L_0$. In the next section we discuss a dS$_4$ mode decomposition
using the highest-weight representations of $\so(4,1)$.

\section{$\so(4,1)$-invariant quasinormal modes}

In this section we describe the $\so(4,1)$-invariant mode
decomposition in terms of (anti-) quasinormal modes. We begin by
defining
\begin{eqnarray}
  G_{\pm}(x;x')&\equiv&G_{E}(x;x')\pm G_{E}(x;x_A')\\
  &=&\frac{1}{8\pi^2}\br{{\frac{1}{1-P(x;x')+i(X^0-X^{'0})\e}}
    \pm\frac{1}{1+P(x;x')+i(X^0+X^{'0})\e}},\nonumber
\end{eqnarray}
where $x_A$ denotes the antipodal point of $x$. These Green functions
fall off like $e^{-2h_\pm t}$ as both arguments are taken to
$\mathcal{I}^+$. Next we introduce `$\O$-modes' as follows:
\begin{equation}
  \Phi^{\pm}_{\O}(x)
  \equiv\frac{\pi}{\sqrt{h_{\pm}}}
  \lim_{t\rightarrow\infty}e^{h_{\pm}t}G_{\pm}(x;\O,t).
\end{equation}
The normalization factor was chosen for future convenience.

In terms of the global coordinates, the $\O$-modes take the explicit
form
\begin{eqnarray}
  \Phi^{-}_{\O}(x)
  &=&\frac{1}{2\pi}\frac{1}{\br{
      \sinh t-i\e-\cosh(t)\cos\Theta_3(\O,x)}}\\
  &=&\frac{1}{2\pi}\frac{1}{\br{
      \sinh{t}-\cosh{t}\cos\Theta_3(\O,x)}}
  -\frac{i}{2\cosh{t}}\delta(\tanh{t}-\cos\Theta_3(\O,x)),
  \nonumber\\
  \Phi^{+}_{\O}(x)
  &=&-\frac{1}{\sqrt{2}\pi}\frac{1}{\br{
      \sinh t-i\e-\cosh(t)\cos\Theta_3(\O,x)}^2}\\
  &=&-\frac{1}{\sqrt{2}\pi}\frac{1}{\br{
      \sinh{t}-\cosh{t}\cos\Theta_3(\O,x)}^2}
  -\frac{i}{\sqrt{2}\cosh^2{t}}\delta'(\tanh{t}-\cos\Theta_3(\O,x)).
  \nonumber
\end{eqnarray}
The delta-functions above are normalized as one-dimensional
delta-functions, that is, such that $\int_{-\infty}^\infty\!dy\
\delta(y)=1$. The $\O$-modes can be expanded in terms of the Euclidean
global $\so(4)$ modes as follows:
\begin{eqnarray}
  \label{eq:omegaeuclidean}
  \Phi_\O^-(x)
  &=&\sqrt{8}\pi\sum\br{\frac{1}{\sqrt{L+1}}Y^*_{Lj}(\O)}\Phi^E_{Lj}(x),\cr
  \Phi_\O^+(x)
  &=&-i\sqrt{16}\pi\sum\br{\sqrt{L+1}Y^*_{Lj}(\O)}\Phi^E_{Lj}(x).
\end{eqnarray}
The lowest-weight and highest-weight modes are respectively given by
\cite{Ng:2012xp}
\begin{equation}
  \Phi^\pm_{lw}(x)\equiv\Phi^\pm_{\O_{SP}}(x),\qquad
  \Phi^\pm_{hw}(x)\equiv\Phi^\pm_{\O_{NP}}(x).
\end{equation}
By construction, the modes $\Phi^{\pm}_{hw}$ are eigenfunctions of
$L_0$ with eigenvalues $-h_{\pm}$ and are annihilated by $M_{-k}$ for
each $k\in\cu{1,2,3}$.  The descendants of the highest-weight modes
are obtained by acting with the $M_{+k}$, for any $k\in\cu{1,2,3}$
(see Appendix \ref{appendix:KV}):
\begin{equation}
  \label{eq:des}
  M_{+K}\Phi_{hw}^{\pm}(x)\equiv M_{+k_1}\cdots M_{+k_n}\Phi^{\pm}_{hw}(x),
\end{equation}
where $K$ is a multi-index denoting the set $\cu{k_1,\ldots,k_n}$.

The southern causal diamond (sometimes called the static patch) is the
intersection of the causal past and future of the south pole. The
highest-weight states are smooth everywhere in this diamond except for
the past horizon where they are singular, and they decay exponentially
towards the future. Therefore they, together with all their
descendants appearing in (\ref{eq:des}) and their complex conjugates,
comprise the quasinormal modes of the southern diamond. The
lowest-weight states (with their descendants and complex conjugates)
are singular on the future horizon and are the antiquasinormal modes
of the southern diamond. To emphasize this we adopt the notation
\begin{equation}
  \Phi^\pm_{QN}(x)\equiv\Phi^\pm_{hw}(x)=\Phi^\pm_{\O_{NP}}(x),\qquad
  \Phi^\pm_{AQN}(x)\equiv\Phi^\pm_{lw}(x)=\Phi^\pm_{\O_{SP}}(x).
\end{equation}
At this point we have eight highest-weight representations of
$\so(4,1)$, with elements
\begin{align}
  \begin{tabular}{llll}
    $M_{+K}\Phi_{QN}^+$,&$M_{+K}\Phi_{QN}^-$,&$M_{+K}\Phi_{QN}^{+*}$,
    &$M_{+K}\Phi_{QN}^{-*}$,\\
    $M_{-K}\Phi_{AQN}^+$,&$M_{-K}\Phi_{AQN}^-$,&$M_{-K}\Phi_{AQN}^{+*}$,
    &$M_{-K}\Phi_{AQN}^{-*}$.
  \end{tabular}
\end{align}
We shall see below that this is an overcomplete set: only the first or
second row of modes is needed to obtain a complete basis.

\section{R-norm}

We wish to expand the scalar field operator in the (anti-)quasinormal
modes. Towards this end it is useful to introduce an inner
product. The Klein-Gordon norms of the $\O$-modes are
\begin{eqnarray}
  \av{\Phi_{\O_1}^\pm,\Phi_{\O_2}^\pm}_{KG}
  &=&\mp\frac{16\pi^2}{h_{\pm}}\Delta_\pm(\O_1,\O_2),\cr
  \av{\Phi_{\O_1}^+,\Phi_{\O_2}^-}_{KG}
  &=&-\av{\Phi_{\O_1}^-,\Phi_{\O_2}^+}_{KG}
  \ =\ \frac{16\pi^2}{\sqrt{2}}\frac{i}{\sqrt{h}}\delta^3(\O_1-\O_2),
\end{eqnarray}
where
\begin{equation}
  \Delta_{\pm}(\O,\O')=\frac{1}{2^{2\mp1}\pi^2}
  \frac{1}{(1-\cos\Theta_3)^{h_{\pm}}}
\end{equation}
denote the two-point functions for a CFT$_3$ operators with dimensions
$h_\pm$. These satisfy
\begin{equation}
  -\int\!d^3\O''\sqrt{h}\ \Delta_+(\O,\O'')\Delta_-(\O'',\O')
  =\frac{1}{\sqrt{h}}\delta^3(\O-\O').
\end{equation}
The norm of a highest-weight quasinormal mode is obtained by setting
$\Omega_1=\Omega_2=\O_{NP}$, which is evidently divergent. Hence the
Klein-Gordon norm is not suitable for quantization of the quasinormal
modes.

Alternate norms have been employed in de Sitter spacetime for a
variety of reasons \cite{Witten:2001kn,Balasubramanian:2002zh,
  Bousso:2001mw,Parikh:2002py,Ng:2012xp}. Here, following
\cite{Ng:2012xp}, a useful `R-norm' can be defined by inserting a
reflection $R$ on $S^3$ across the equator:
\begin{eqnarray}
  R:(\psi,\theta,\phi)&\rightarrow&(\pi-\psi,\theta,\phi),\cr
  \rn{\Phi_1,\Phi_2}&\equiv&\av{\Phi_1,R\Phi_2}_{KG}.
\end{eqnarray}
With respect to this R-norm,
\begin{equation}
  \rn{\Phi_{\O_1}^\pm,\Phi_{\O_2}^\pm}
  =\mp\frac{16\pi^2}{h_{\pm}}\Delta_\pm(\O_1,R\O_2).
\end{equation}
In particular, the norms of the highest-weight quasinormal modes are
simply
\begin{equation}
  \rn{\Phi_{QN}^{+},\Phi_{QN}^{+}}=-1,\qquad
  \rn{\Phi_{QN}^-,\Phi_{QN}^-}=1,\qquad
  \rn{\Phi_{QN}^+,\Phi_{QN}^-}=0,
\end{equation}
while the R-inner product between a quasinormal mode and the complex
conjugate of any quasinormal mode vanishes.

Changing the norm affects the hermiticity properties of the 10 real
Killing vector fields which generate the dS$_4$ isometries. Under the
Klein-Gordon norm, their adjoints are
\begin{equation}
  \av{L_0f,g}_{KG}=\av{f,-L_0g}_{KG},\quad
  \av{J_kf,g}_{KG}=\av{f,-J_kg}_{KG},\quad
  \av{M_{\mp k}f,g}_{KG}=\av{f,-M_{\mp k}g}_{KG},
\end{equation}
so that the Killing generators are all antihermitian. However, under
the modified R-norm,
\begin{equation}
  \label{eq:algebra}
  \rn{L_0f,g}=\rn{f,L_0g},\qquad
  \rn{J_kf,g}=\rn{f,-J_k g},\qquad
  \rn{M_{\mp k}f,g}=\rn{f,M_{\pm k}g}.
\end{equation}
To recover antihermitian generators in the R-norm, one must send
$M_k+M_{-k}\rightarrow i(M_k+M_{-k})$ and $L_0\rightarrow iL_0$ while
keeping the rest of the generators the same. The Lie bracket algebra
of the antihermitian vector fields is then $\so(3,2)$ rather than
$\so(4,1)$. See Appendix \ref{appendix:KV} for more details.

Interestingly $\so(3,2)$ is the symmetry group of a CFT in $2+1$
dimensions. This suggests that the quantum states on which these
generators act could belong to a $2+1$-dimensional CFT, which fits in
nicely with the dS$_4$/CFT$_3$ conjecture.

Using \eqref{eq:algebra} we can compute the norm of the
descendants. For example, the norm of the first descendant is (not
summing over $k$)
\begin{equation}
  \rn{M_{+k}\Phi^{\pm}_{QN},M_{+k}\Phi^{\pm}_{QN}}
  =\rn{\Phi^{\pm}_{QN},M_{-k}M_{+k}\Phi^{\pm}_{QN}}
  =2h_{\pm}\rn{\Phi^{\pm}_{QN},\Phi^{\pm}_{QN}}.
\end{equation}
Observe that under this R-norm, the descendants of $\Phi^+_{QN}$ are
orthogonal to those of $\Phi^-_{QN}$. For the $\so(3)$-symmetric
states, we provide the exact formula in Appendix \ref{appendix:Norm}.

\section{Completeness of quasinormal modes}

In this section we show that the quasinormal modes
\begin{equation}
  \label{eq:completeset}
  \cu{M_{+K}\Phi_{QN}^-,\qquad M_{+K}\Phi_{QN}^{-*},\qquad
    M_{+K}\Phi_{QN}^+,\qquad M_{+K}\Phi_{QN}^{+*}}
\end{equation}
form a complete set in the sense that the Euclidean Green function can
be written as a simple sum over such modes. In particular, the
antiquasinormal modes are not needed.

First we note from \eqref{eq:omegaeuclidean} that the quasinormal
modes can be written as linear combinations of the global Euclidean
modes, without using their complex conjugates. Therefore they are
themselves Euclidean modes, and the Euclidean vacuum obeys
\begin{equation}
  \label{eq:euclideanvaccuum}
  \av{M_{+K}\Phi^\pm_{QN},\hat\Phi}_{R}\ket{0_E}=0.
\end{equation}
Note that this relation, unlike the corresponding one for the global
Euclidean modes, is manifestly dS-invariant because the quasinormal
modes lie in representations of $\so(4,1)$.

Let us now assume that we can expand the field operator in the
presumably complete basis \eqref{eq:completeset}:
\begin{align}
  \hat\Phi=\sum_{K,K'}\Big(
  &N_{KK'}^+\rn{M_{+K}\Phi_{QN}^+,\hat\Phi}M_{+K'}\Phi_{QN}^+
  -N_{K'K}^+\rn{M_{+K}\Phi_{QN}^{+*},\hat\Phi}M_{+K'}\Phi_{QN}^{+*}\\
  &+N_{KK'}^-\rn{M_{+K}\Phi_{QN}^-,\hat\Phi}M_{+K'}\Phi_{QN}^-
  -N_{K'K}^-\rn{M_{+K}\Phi_{QN}^{-*},\hat\Phi}M_{+K'}\Phi_{QN}^{-*}
  \Big)\nonumber,
\end{align}
where the $N^{\pm}_{KK'}$ are defined through
\begin{equation}
  \sum_{K'}N^{\pm}_{KK'}\rn{M_{+K'}\Phi_{QN}^\pm,M_{+L}\Phi_{QN}^\pm}
  =\delta_{KL}.
\end{equation}
Then, using \eqref{eq:euclideanvaccuum}, the quasinormal mode Green
function is given by
\begin{equation}
  \label{eq:sumovermodes}
  G(x;x')
  =\sum_{K,K'}\Phi^{+}_K(x)\Phi_{K'}^{+*}(Rx')N^{+}_{KK'}
  +\sum_{K,K'}\Phi^{-}_K(x)\Phi_{K'}^{-*}(Rx')N^{-}_{KK'},
\end{equation}
where $\Phi^{\pm}_K\equiv M_{+K}\Phi_{QN}^\pm$.

A demonstration that the function $G(x;x')$ so obtained is indeed the
standard Euclidean Green function $G_E(x;x')$ implies that the
quasinormal modes in \eqref{eq:completeset} form a complete basis, in
the sense that they satisfy
\begin{eqnarray}
  i\delta^3(\O-\O')
  &=&\sqrt{\gamma}n^\mu\sum_{K,K'}N^+_{KK'}
  \br{\Phi^{+*}_K(t,\O)\nabla_\mu\Phi^+_{K'}(t,R\O')
    -\Phi^+_K(t,\O)\nabla_\mu\Phi^{+*}_{K'}(t,R\O')}\cr
  &&+\pa{+\leftrightarrow-}\\
  0&=&\sum_{K,K'}N^+_{KK'}\br{\Phi^+_K(t,\O)\Phi^{+*}_{K'}(t,R\O')
    -\Phi^{+*}_K(t,\O)\Phi^+_{K'}(t,R\O')}+\pa{+\leftrightarrow-},
  \nonumber
\end{eqnarray}
on a constant time slice with normal vector $n^\mu$ and induced metric
$\gamma_{\mu\nu}$. Indeed, these two equations can be used to
construct a retarded Green function, which in turn provides a solution
to the wave equation with arbitrary initial data. Hence any suitably
smooth solution to the wave equation can be decomposed on a Cauchy
surface in terms of such a set of modes.

First, we would like to evaluate the sum \eqref{eq:sumovermodes} for
the case $(x;x')=(t,\O_{SP};t',\O_{SP})$ where both points lie on the
south pole observer's worldline. The functions $\Phi^\pm_K(t,\O_{SP})$
are nonzero only for spherically symmetric descendants
$L_{+1}^n\Phi^\pm_{QN}(t,\O)$ where
$L_{\mp1}\equiv\displaystyle\sum_{k=1}^3M_{\mp k}M_{\mp k}.$ The norm
for such states is calculated in Appendix \ref{appendix:Norm} and is
given by
\begin{equation}
  \rn{L_{+1}^n\Phi_{QN}^{\pm},L_{+1}^m\Phi^{\pm}_{QN}}
  =\frac{\Gamma(2+2n)\Gamma(2h_{\pm}+2n-1)}{\Gamma(2h_{\pm}-1)}
  \rn{\Phi_{QN}^{\pm},\Phi_{QN}^{\pm}}\delta_{nm},
\end{equation}
while the modes at $\O=\O_{SP}$ are given by
\begin{eqnarray}
  \label{eq:descendants}
  L^n_{+1}\Phi_{QN}^{-}(t,\O_{SP})&=&\frac{\Gamma(2n+2)}{2\pi}
  \frac{e^{-nt}}{\left(e^{+t}-i\e\right)^{n+1}},\cr
  L^n_{+1}\Phi_{QN}^{+}(t,\O_{SP})&=&
  -\frac{\Gamma(2n+3)}{2\sqrt{2}\pi}
  \frac{e^{-nt}}{\left(e^{+t}-i\e\right)^{n+2}}.
\end{eqnarray}
Using
\begin{eqnarray}
  \Big(L^n_{+1}\Phi_{QN}^{-}(t,R\O_{SP})\Big)^{*}
  &=&-L^n_{+1}\Phi_{QN}^{-}(-t,\O_{SP})\cr
  \Big(L^n_{+1}\Phi_{QN}^{+}(t,R\O_{SP})\Big)^{*}
  &=&L^n_{+1}\Phi_{QN}^{+}(-t,\O_{SP}),
\end{eqnarray}
the full sum \eqref{eq:sumovermodes} is
\begin{eqnarray}
  G(t,\O_{SP};t',\O_{SP})
  &=&-\frac{1}{4\pi^2}\sum_{k=0}^\infty
  \cu{\frac{(2k+1)e^{-k(t-t')}}{\br{\pa{e^{+t}-i\e}\pa{e^{-t'}-i\e}}^{k+1}}
    +\frac{(2k+2)e^{-k(t-t')}}{\br{\pa{e^{+t}-i\e}\pa{e^{-t'}-i\e}}^{k+2}}}\cr
  &=&-\frac{1}{16\pi^2}\frac{1}{\sinh^2[(t-t')/2]
    -i\e\tilde{s}(x;x')},
\end{eqnarray}
where
\begin{equation}
  \tilde{s}(x;x')\equiv\frac{\sinh{t}-\sinh{t'}}{1+e^{t'-t}}.
\end{equation}
Noting that for small $\e$, $\tilde{s}(x;x')$ is equivalent to
$s(x;x')$ defined in \eqref{eq:ss}, it follows that this Green
function agrees with that in \eqref{eq:GE} on the south pole
observer's worldline. Since the construction of our Green function is
dS-invariant\footnote{This follows from the fact that the Green
  function is just a position-space representation of the projection
  operator onto the highest-weight representation of the
  three-dimensional conformal group characterized by the
  highest-weight $-h$, as can be seen by writing out this projection
  as a sum over complete states of the representation and using the
  definition of $\so(4,1)$ generators.}, agreement on this worldline
implies that this Green function equals the Euclidean one on any two
{\it timelike} separated points.

For {\it spacelike} separated points, we find from
\eqref{eq:sumovermodes} that
\begin{equation}
  G(t,\O_{SP};t',\O_{NP})
  =\frac{1}{8\pi^2\br{1+\cosh{(t+t')}}}
  =G_E(t,\O_{SP};t',\O_{NP}).
\end{equation}
By dS-invariance, we can extend this to any two spacelike-separated
points. This concludes the proof that the quasinormal Green function
\eqref{eq:sumovermodes} is indeed the Euclidean Green function.

\section{Results for general light scalars $(m^2 \ell^2<9/4)$}

In the general case of a light scalar with $m^2\ell^2<9/4$, we can
write out the explicit form of the Euclidean two-point function as
(see for instance \cite{Bousso:2001mw})
\begin{equation}
  \label{eq:massiveGE}
  G_E(x;x')
  =\frac{\Gamma(h_+)\Gamma(h_-)}{16\pi^2}
  F\!\br{h_+,h_-,2,\frac{1+P(x;x')-is(x;x')\e }{2}},
\end{equation}
where
\begin{equation}
  h_\pm=\frac{3}{2}\pm\mu,\qquad
  \mu=\sqrt{\frac{9}{4}-m^2\ell^2}.
\end{equation}
The asymptotic behaviors of the Euclidean Green function are:
\begin{eqnarray}
  \lim_{t'\rightarrow\infty}G_E(t,\O;t',\O_{NP})
  &=&\frac{\Gamma(h_--h_+)\Gamma(h_+)}{2^{4-2h_+}\pi^2 \Gamma(2-h_+)}
  \frac{e^{-h_+ t'}}{(\sinh{t}-i\e+\cosh{t}\cos{\psi})^{h_+}}
  +(h_+\leftrightarrow h_-),\cr
  \lim_{t'\rightarrow\infty}G_E(t,\O;-t',\O_{SP})
  &=&e^{-i\pi h_+}\frac{\Gamma(h_--h_+)\Gamma(h_+)}{2^{4-2h_+}\pi^2 \Gamma(2-h_+)}
  \frac{e^{-h_+ t'}}{(\sinh{t}-i\e+\cosh{t}\cos{\psi})^{h_+}}\cr
  &&+(h_+\leftrightarrow h_-).
\end{eqnarray}
Note that in dealing with the branch-cut of $G_E(t,\O;t',\O')$, we go
under (above) it when $t>t'$ ($t<t'$) in accordance with the
$i\e$-prescription.

Let us define $G_\pm$ as
\begin{equation}
  \label{eq:Gpm}
  G_\pm(x;x')\equiv G_E(x;x')-e^{i\pi h_\mp}G_E(x;x_A').
\end{equation}
These satisfy the future boundary conditions in
Ref.~\cite{Anninos:2011jp} in the region $P<-1$. Now, we define the
highest-weight modes as
\begin{equation}
  \Phi^{\pm}_{QN}(x)
  \equiv\lim_{t'\rightarrow\infty}e^{h_{\pm} t'}G_{\pm}(t,\O;t',\O_{NP}).
\end{equation}
The $\Phi_{QN}^\pm$ are explicitly given by
\begin{eqnarray}
  \label{eq:qnpm}
  \Phi^\pm_{QN}(x)
  &=&\frac{1}{4\pi^{5/2}}
  \frac{\Gamma(\mp\mu)\Gamma(h_\pm)\pa{1-e^{\mp2\pi i\mu}}}
  {\br{\sinh{t}-i\e+\cosh{t}\cos{\psi}}^{h_\pm}}
  \ =\ \frac{1}{4\pi^{5/2}}
  \frac{\Gamma(\mp\mu)\Gamma(h_\pm)\pa{1-e^{\mp2\pi i\mu}}}
  {\cosh^{h_\pm}{t}\br{\tanh(t-i\e)+\cos{\psi}}^{h_\pm}}.
  \nonumber\\
\end{eqnarray}
The asymptotic behavior of the modes as $t\rightarrow\infty$ is
\begin{equation}
  \lim_{t\rightarrow \infty}\Phi_{QN}^{\pm}(t,\Omega)
  =\frac{2^{3-h_{\pm}}}{\sqrt{\pi}}
  \Gamma(\mp\mu)\Gamma(h_\pm)\pa{1-e^{\mp2\pi i\mu}}
  \Delta_\pm(\Omega,\Omega_{NP})e^{-h_{\pm}t}
  \mp\frac{4i}{\mu}
  \frac{\delta^3(\Omega-\Omega_{NP})}{\sqrt{h}}e^{-h_\mp t}.
\end{equation}
We have defined\footnote{Here, a useful identity on $S^3$ is
  \begin{equation}
    \br{1-\cos{\Theta_3(\O,\O')}}^{-h}
    =2^{2+h}\pi\sin{(\pi h)}\Gamma(2-2h)
    \sum\frac{\Gamma(L+h)}{\Gamma(3+L-h)}Y_{Lj}(\O)Y^*_{Lj}(\O').
  \end{equation}}
\begin{eqnarray}
  \Delta_\pm(\O,\O')
  &=&\frac{2^{3(h_\pm-1)}}{\pi}\Gamma(2-2h_\pm)\sin{(h_\pm\pi)}
  \sum\frac{\Gamma(h_\pm+L)}{\Gamma(h_\mp+L)}Y_{Lj}(\O)Y^*_{Lj}(\O')\cr
  &=&\frac{1}{2^{5-2h_\pm}\pi^2}
  \frac{1}{\br{1-\cos{\Theta_3(\O,\O')}}^{h_\pm}},
\end{eqnarray}
which satisfy
\begin{equation}
  \frac{\pi^2}{8\cos^2{(\pi \mu)}\Gamma(2-2h_+)\Gamma(2-2h_-)}
  \int\!d^3\O''\sqrt{h}\ \Delta_+(\O,\O'')\Delta_-(\O'',\O')
  =\frac{1}{\sqrt{h}}\delta^3(\O-\O').
\end{equation}
The norm is easily evaluated at $\mathcal{I}^+$ to be
\begin{eqnarray}
  \av{\Phi_{\O_1}^\pm,\Phi_{\O_2}^\pm}_R
  &=&\frac{2^{5-h_\pm}}{\sqrt{\pi}}\Gamma(\mp\mu)\Gamma(h_\pm)
  \sin^2(\pi\mu)\Delta_\pm(\O_1,R\O_2),\cr
  \av{\Phi_{\O_1}^\pm,\Phi_{\O_2}^\mp}_R
  &=&\pm\frac{4i}{\mu}\pa{1-e^{\pm2\pi i\mu}}
  \frac{\delta^3(\Omega_1-R\Omega_2)}{\sqrt{h}}.
\end{eqnarray}
As such, we find that the R-norms of the quasinormal modes are
\begin{equation}
  \rn{\Phi^\pm_{QN},\Phi^\pm_{QN}}
  =\Gamma(\mp\mu)\Gamma(h_{\pm})\frac{\sin^2(\pi\mu)}{\pi^{5/2}},\qquad
  \rn{\Phi^+_{QN},\Phi^-_{QN}}=0.
\end{equation}
The rest of the discussion on the induced norms of the descendants
carries over from the $m^2\ell^2=2$ case.

Next, we follow our previous strategy of showing that the mode sum and
the Euclidean Green function agree on the south pole observer's
worldline. Again, we evaluate
\begin{eqnarray}
  L_{+1}^n\Phi_{QN}^\pm(t,\O_{SP})&=&
  \frac{\Gamma(\mp\mu)(1-e^{\mp2\pi i \mu})}{4\pi^{5/2} }
  \frac{\Gamma(2n+3)\Gamma(h_\pm +n)}{2\Gamma(n+2)}
  \frac{e^{-n t}}{(e^t-i\e)^{n+h_{\pm}}}.
\end{eqnarray}
As before, the norm for such states is
\begin{equation}
  \rn{L_{+1}^n\Phi_{QN}^\pm,L_{+1}^n\Phi^\pm_{QN}}
  =\frac{\Gamma(2+2n)\Gamma(2h_\pm+2n-1)}{\Gamma(2h_\pm-1)}
  \rn{\Phi_{QN}^\pm,\Phi_{QN}^\pm}.
\end{equation}
Note that
\begin{equation}
  \Big(L^n_{+1}\Phi_{QN}^{\pm }(t,R\O_{SP})\Big)^{*}
  =e^{i\pi h_{\pm}}L^n_{+1}\Phi_{QN}^{\pm }(-t,\O_{SP}).
\end{equation}
The quasinormal mode Green function \eqref{eq:sumovermodes} is then
\begin{equation}
  \label{eq:greensummassive}
  G(t,\O_{SP};t',\O_{SP})
  =\sum_{n=0}^\infty\br{
    \frac{c^+_ne^{-n(t-t')}}
    {\br{\pa{e^t-i\e}\pa{e^{-t'}-i\e}}^{n+h_+}}
    +\frac{c_n^-e^{-n(t-t')}}
    {\br{\pa{e^t-i\e}\pa{e^{-t'}-i\e}}^{n+h_-}}}
\end{equation}
where
\begin{equation}
  \label{eq:coefficients}
  c^\pm_n=-\frac{e^{-i\pi h_\pm}}{2\pi^2\sin{(\pm\pi\mu)}}
  \frac{\Gamma(\frac{3}{2}+n)\Gamma(h_\pm+n)}
  {\Gamma(1+n\pm\mu)\Gamma(1+n)}.
\end{equation}
In Appendix \ref{appendix:Green}, we show that on the south pole
observer's worldline, this Green function is equal to the Euclidean
Green function. Thus by dS-invariance of both Green functions, they
agree for any two timelike separated points in dS$_4$.

For spacelike separated points, we consider the Euclidean Green
function with one point at the south pole and the other point at the
north pole. One notices that
\begin{equation}
  G_E(t,\O_{SP};t',\O_{NP})=G_E(t,\O_{SP};t',\O_{SP})|_{t'\rightarrow -t'+i\pi}.
\end{equation}
\clearpage
Since these points are spacelike separated, we do not have to worry
about the $i\e$-prescription and the Green function is real. If we had
evaluated the quasinormal mode Green function
$G(t,\O_{SP};t',\O_{NP})$, then we would have obtained the same sum as
in \eqref{eq:greensummassive}, provided that we sent
$t'\rightarrow-t'$ and removed the phase $e^{-i\pi h_\pm}$ from the
coefficients \eqref{eq:coefficients}. This is equivalent to sending
$t'\rightarrow-t'+i\pi$ and hence by dS-invariance we have proved that
for any two {\it spacelike} separated points, the quasinormal mode
Green function is the Euclidean Green function.

\section{Southern modes and T-norm}
In this section we find quasinormal modes that vanish in the northern
or southern diamonds -- the analogues of Rindler modes in Minkowski
space. We begin with the expression \eqref{eq:qnpm} for the
lowest-weight mode
\begin{equation}
  \Phi^\pm_{QN}(x)=\frac{1}{4\pi^{5/2}}
  \frac{\Gamma(\mp\mu)\Gamma(h_\pm)\pa{1-e^{\mp2\pi i\mu}}}
  {\br{\sinh{t}-i\e+\cosh{t}\cos{\psi}}^{h_\pm}}.
\end{equation}
For generic mass $\Phi^{\pm}_{QN}(x)$ has a branch cut on the past
horizon of the southern observer at $\tanh{t}=-\cos\psi$. We have
chosen the phase convention so that the denominator is real above the
past horizon. Crossing the past horizon gives an extra phase of
$e^{i\pi h_\pm}$. It follows that the southern mode
\begin{equation}
  \Phi^\pm_{QN,S}(x)\equiv\Phi^{\pm}_{QN}(x)+\Phi^{\pm*}_{QN}(x)
\end{equation}
vanishes below the past horizon. Similarly the northern mode
\begin{equation}
  \Phi^\pm_{QN,N}(x)\equiv
  e^{i\pi h_\pm}\Phi^\pm_{QN}(x)+e^{-i\pi h_\pm}\Phi^{\pm*}_{QN}(x).
\end{equation}
vanishes above the past horizon. The R-norms between these modes are
\begin{eqnarray}
  \rn{\Phi^\pm_{QN,S},\Phi^\pm_{QN,S}}
  &=&\rn{\Phi^\pm_{QN,N},\Phi^\pm_{QN,N}}
  \ =\ \rn{\Phi^\pm_{QN,N},\Phi^\mp_{QN,N}}
  \ =\ \rn{\Phi^\mp_{QN,S},\Phi^\mp_{QN,S}}\ =\ 0,\nonumber\\
  \rn{\Phi^\pm_{QN,N},\Phi^\pm_{QN,S}}
  &=&-2i\sin(\pi h_\pm)\rn{\Phi^\pm_{QN},\Phi^\pm_{QN}}
  \ =\ -2i\sin(\pi h_\pm)\Gamma(\mp\mu)\Gamma(h_{\pm})
  \frac{\sin^2(\pi\mu)}{\pi^{5/2}}.\nonumber\\
\end{eqnarray}
On the other hand, the R-norm for the global quasinormal modes is
closely related to time-reflection:
\begin{equation}
  \rn{f,\Phi_K^\pm}=e^{-i\pi h_{\pm}}\av{f,T\Phi_K^{\pm*}}_{KG},
  \qquad T:t\rightarrow-t.
\end{equation}
While the R-norm has no analogue in the static patch, the T-norm is
easily generalizable to the southern diamond as
\begin{equation}
  \av{\Phi_{K}^\pm,\Phi_{K'}^{\pm}}_{T,B_S^3}
  \equiv\av{T\Phi^{\pm*}_{K},\Phi_{K'}^\pm}_{KG,B_S^3},
\end{equation}
where $B_S^3$ denotes the integral over a complete slice in the
southern diamond. We have
\begin{eqnarray}
  \av{\Phi_{QN,S}^\pm,T\Phi_{QN,S}^\pm}_{T,B_S^3}
  &=&\av{\Phi_{QN,S}^\pm,T\Phi_{QN,S}^\mp}_{T,B_S^3}
  \ =\ \av{\Phi_{QN,S}^\pm,\Phi_{QN,S}^\mp}_{T,B_S^3}\ =\ 0,\cr
  \av{\Phi^\pm_{QN,S},\Phi^\pm_{QN,S}}_{T,B_S^3}
  &=&2i\sin(\pi h_\pm)\Gamma(\mp\mu)\Gamma(h_{\pm})
  \frac{\sin^2(\pi\mu)}{\pi^{5/2}}.
\end{eqnarray}

\vspace{.5cm}

\section*{Acknowledgements}
It has been a great pleasure discussing this work with Tatsuo
Azeyanagi and Dionysios Anninos. This work was supported in part by
DOE grant DE-FG02-91ER40654 and the Fundamental Laws Initiative at
Harvard.

\vspace{1cm}

\appendix

\section{Appendix: dS$_4$ Killing vectors}
\label{appendix:KV}

In global coordinates, the 10 Killing vectors of dS$_4$ are given by:
\begin{eqnarray}
  L_0&=&\cos{\psi}\pd_t-\tanh{t}\sin{\psi}\pd_\psi,\cr
  M_{\mp 1}&=&\pm\sin{\psi}\sin{\theta}\sin{\phi}\pd_t
  +(1\pm\tanh{t}\cos{\psi})\sin{\theta}\sin{\phi}\pd_\psi\cr
  & &+(\cot{\psi}\pm\tanh{t}\csc{\psi})
  (\cos{\theta}\sin{\phi}\pd_\theta+\csc{\theta}\cos{\phi}\pd_\phi),\cr
  M_{\mp2}&=&\pm\sin{\psi}\sin{\theta}\cos{\phi}\pd_t
  +(1\pm\tanh{t}\cos{\psi})\sin{\theta}\cos{\phi}\pd_\psi\cr
  & &+(\cot{\psi}\pm\tanh{t}\csc{\psi})
  (\cos{\theta}\cos{\phi}\pd_\theta-\csc{\theta}\sin{\phi}\pd_\phi),\cr
  M_{\mp3}&=&\pm\sin{\psi}\cos{\theta}\pd_t
  +(1\pm\tanh{t}\cos{\psi})\cos{\theta}\pd_\psi
  -(\cot{\psi}\pm\tanh{t}\csc{\psi})\sin{\theta}\pd_\theta,\cr
  J_1&=&\cos{\phi}\pd_\theta-\sin{\phi}\cot{\theta}\pd_\phi,\cr
  J_2&=&-\sin{\phi}\pd_\theta-\cos{\phi}\cot{\theta}\pd_\phi,\cr
  J_3&=&\pd_\phi.
\end{eqnarray}
Their non-zero commutators are:
\begin{eqnarray}
  \br{J_i,J_j}&=&\sum_{k=1}^3\e_{ijk}J_k,\qquad
  \br{J_i,M_{\pm j}}\ \
  =\ \ \sum_{k=1}^3\e_{ijk}M_{\pm k},\cr
  \br{L_0,M_{\pm i}}&=&\mp M_{\pm i},\qquad\,\
  \br{M_{+i},M_{-j}}\ \
  =\ \ 2L_0\delta_{ij}+2\sum_{k=1}^3\e_{ijk}J_k.
\end{eqnarray}
As expected, these are the relations which define the $\so(4,1)$
algebra. The commutators on the first line indicate that the $J_i$
generate an $\so(3)$ subalgebra, under which the $M_{+i}$ and the
$M_{-i}$ transform as vectors. The second line implies that for each
$i\in\cu{1,2,3}$, the Killing vectors $M_{\pm i}$ and $L_0$ form an
$\so(2,1)$ subalgebra satisfying (not summing over $i$)
\begin{equation}
  \br{M_{+i},M_{-i}}=2L_0,\qquad
  \br{L_0,M_{\pm i}}=\mp M_{\pm i}.
\end{equation}
The scalar Laplacian is a Casimir operator. It reads:
\begin{equation}
  \ell^2\nabla^2=-L_0(L_0-3)+\sum_{i=1}^3M_{-i} M_{+i}+J^2.
\end{equation}
The convention is that $J^2=-L(L+1)$ on the spherical harmonics
$Y_{Lj}$. The conformal Killing vectors of the $S^3$ are given by the
restriction of dS$_4$ Killing vectors on $\mathcal{I}^+$:
\begin{eqnarray}
  L_0&=&-\sin{\psi}\pd_\psi,\cr
  M_{\mp 1}&=&(1\pm\cos{\psi})\sin{\theta}\sin{\phi}\pd_\psi
  +(\cot{\psi}\pm\csc{\psi})
  (\cos{\theta}\sin{\phi}\pd_\theta+\csc{\theta}\cos{\phi}\pd_\phi),\cr
  M_{\mp2}&=&(1\pm\cos{\psi})\sin{\theta}\cos{\phi}\pd_\psi
  +(\cot{\psi}\pm\csc{\psi})
  (\cos{\theta}\cos{\phi}\pd_\theta-\csc{\theta}\sin{\phi}\pd_\phi),\cr
  M_{\mp3}&=&(1\pm\cos{\psi})\cos{\theta}\pd_\psi
  -(\cot{\psi}\pm\csc{\psi})\sin{\theta}\pd_\theta,\cr
  J_1&=&\cos{\phi}\pd_\theta-\sin{\phi}\cot{\theta}\pd_\phi,\cr
  J_2&=&-\sin{\phi}\pd_\theta -\cos{\phi}\cot{\theta}\pd_\phi,\cr
  J_3&=&\pd_\phi.
\end{eqnarray}
To relate the above de Sitter generators to the embedding coordinates
$X$ defined by
\begin{equation}
  \eta_{\mu\nu}X^\mu X^\nu=\ell^2,
\end{equation}
where $\eta$ has signature $(4,1)$ and the usual Lorentz generators
are given by
\begin{equation}
  M_{\mu\nu}=X_\mu\pd_\nu-X_\nu\pd_\mu,
\end{equation}
with commutators
\begin{equation}
  \label{eq:SO41com}
  \br{M_{\mu\nu},M_{\alpha\beta}}
  =\eta_{\alpha\nu}M_{\mu\beta}-\eta_{\alpha\mu}M_{\nu\beta}
  -\eta_{\beta\nu}M_{\mu\alpha}+\eta_{\beta\mu}M_{\nu\alpha},
\end{equation}
we have for $i,j,k\in\cu{1,2,3}$
\begin{eqnarray}
  L_0&=&M_{40},\cr
  M_{\mp k}&=&M_{4k}\mp M_{0k},\cr
  J_i&=&-\e_{ijk}M_{jk}.
\end{eqnarray}
The standard Klein-Gordon adjoint acts as:
\begin{equation}
  M_{\mu\nu}^\dagger=-M_{\mu\nu},
\end{equation}
where the adjoint is defined in the standard way as
$\av{f,M^\dagger g}_{KG}\equiv\av{Mf,g}_{KG}$. The action of $R$ on the
Killing vectors is:
\begin{equation}
   L_0\rightarrow-L_0,\qquad J_k\rightarrow J_k,\qquad
   M_{\pm k}\rightarrow-M_{\mp k},
\end{equation}
or equivalently, for $j,k\in\cu{1,2,3}$,
\begin{equation}
  M_{40}\rightarrow-M_{40},\qquad M_{4k}\rightarrow-M_{4k},\qquad
  M_{0k}\rightarrow M_{0k},\qquad M_{jk}\rightarrow M_{jk}.
\end{equation}
In the R-norm, if we define $M^{\dagger_R}$ as
$\rn{f,M^{\dagger_R}g}\equiv\rn{Mf,g}$ then
\begin{equation}
  M_{40}^{\dagger_R}=M_{40},\qquad
  M_{4k}^{\dagger_R}=M_{4k},\qquad
  M_{0k}^{\dagger_R}=-M_{0k},\qquad
  M_{jk}^{\dagger_R}=-M_{jk}.
\end{equation}
With respect to the R-norm, the antihermitian generators are
$iM_{40}$, $iM_{4k}$, $M_{0k}$ and $M_{jk}$. On the other hand, we
have from \eqref{eq:SO41com} that for $i,j,k\in\cu{1,2,3}$,
\begin{align}
  \begin{tabular}{lll}
    $\br{M_{0i},M_{0j}}=-\eta_{00}M_{ij}$,&
    $\br{M_{0i},M_{04}}=-\eta_{00}M_{i4}$,&
    $\br{M_{0i},M_{j4}}=\eta_{ij}M_{04}$,\\
    $\br{M_{0i},M_{jk}}= \eta_{ij}M_{0k}$,&
    $\br{M_{04},M_{j4}}=-\eta_{44}M_{0j}$,&
    $\br{M_{j4},M_{k4}}=-\eta_{44}M_{jk}$,
  \end{tabular}
\end{align}
while the $M_{jk}$ obey the usual $\so(3)$ algebra. Now, if we sent
$M_{40}\rightarrow iM_{40}$ and $M_{4k}\rightarrow i M_{4k}$, we would
get the same algebra but with $\eta_{44}\rightarrow-\eta_{44}$. This
demonstrates that the insertion of $R$ in the norm transforms
$\so(4,1)$ into $\so(3,2)$.

\section{Appendix: Norm for spherically symmetric states}
\label{appendix:Norm}

Consider the operator $L_{\mp1}\equiv\sum\limits_{k=1}^3M_{\mp k}M_{\mp k}$,
which evidently satisfies $\br{J_k,L_{\mp1}}=0$. Defining
$\ket{h+n}\equiv L_{+1}^n\ket{h}$, where $\ket{h}$ is the spherically
symmetric highest-weight state with $J^2\ket{h}=0$ and
$L_0\ket{h}=-h\ket{h}$, we have
\begin{equation}
  \br{L_{+1},L_{-1}}\ket{h+n}=4L_0\pa{2L_0^2+2\nabla^2-3}\ket{h+n}.
\end{equation}
The Casimir is
\begin{equation}
  \nabla^2
  =-L_0(L_0-3)+M_{-k}M_{+k}+J^2
  =-L_0(L_0+3)+M_{+k}M_{-k}+J^2
\end{equation}
and $\nabla^2\ket{h+n}= -h(h-3)\ket{h+n}$. Then using
\begin{equation}
  \br{L_{-1},L_{+1}}\ket{h+n}
  =4(h+2n)(8n^2+8nh+6h-3)\ket{h+n},
\end{equation}
it is straightforward to show that
\begin{eqnarray}
  \bra{h}L^n_{-1}L^n_{+1}\ket{h}
  &=&4n(n+h-1)(2n+1)(2n+2h-3)\bra{h}L^{n-1}_{-1}L^{n-1}_{+1}\ket{h}\cr
  &=&\frac{\Gamma(2+2n)\Gamma(2h+2n-1)}{\Gamma(2h-1)}\av{h|h}.
\end{eqnarray}

\section{Appendix: Green function at the south pole}
\label{appendix:Green}

We wish to evaluate the sum \eqref{eq:sumovermodes} over the
quasinormal modes for the massive case,
\begin{eqnarray}
  G(t,\Omega_{SP};t',\Omega_{SP})
  &=&\br{\pa{e^t-i\e}\pa{e^{-t'}-i\e}}^{-h_+}\sum_nc_{n}^+
  \br{\frac{e^{-(t-t')}}{\pa{e^t-i\e}\pa{e^{-t'}-i\e}}}^n\cr
  &&+\br{\pa{e^t-i\e}\pa{e^{-t'}-i\e}}^{-h_-}\sum_nc_{n}^-
  \br{\frac{e^{-(t-t')}}{\pa{e^t-i\e}\pa{e^{-t'}-i\e}}}^n
\end{eqnarray}
where
\begin{equation}
  c^\pm_n=\pm\frac{e^{-i\pi h_\pm}}{(-2\pi^2)\sin{(\pi\mu)}}
  \frac{\Gamma(\frac{3}{2}+n)\Gamma(h_\pm+n)}
  {\Gamma(1+n\pm\mu)\Gamma(1+n)}.
\end{equation}
Each sum combines into a hypergeometric function with argument shifted
by $\e$
\begin{eqnarray}
  &&2\sin(\pi\mu)G(t,\Omega_{SP};t',\Omega_{SP})\cr
  &=&\frac{e^{-i\pi h_+}}{\br{\pa{e^t-i\e}\pa{e^{-t'}-i\e}}^{h_+}}
  \frac{\Gamma\pa{\frac{3}{2}}\Gamma(h_+)}{(-\pi^2)\Gamma(1+\mu)}
  F\!\br{h_+,\frac{3}{2};1+\mu;
    \frac{e^{-(t-t')}}{\pa{e^t-i\e}\pa{e^{-t'}-i\e}}}\cr
  &-&\frac{e^{-i\pi h_-}}{\br{\pa{e^t-i\e}\pa{e^{-t'}-i\e}}^{h_-}}
  \frac{\Gamma\pa{\frac{3}{2}}\Gamma(h_-)}{(-\pi^2)\Gamma(1-\mu)}
  F\!\br{h_-,\frac{3}{2};1-\mu;
    \frac{e^{-(t-t')}}{\pa{e^t-i\e}\pa{e^{-t'}-i\e}}}.
\end{eqnarray}
Kummer's quadratic transformation
\begin{equation}
  F\!\br{\alpha,\beta,2\beta,\frac{4z}{(1+z)^2}}
  =(1+z)^{2\alpha}F\!\br{\alpha,\alpha-\beta+\frac{1}{2},
    \beta+\frac{1}{2},z^2}
\end{equation}
with
\begin{equation}
  z\equiv\frac{e^{-(t-t')/2}}{\br{\pa{e^t-i\e}\pa{e^{-t'}-i\e}}^{1/2}}
\end{equation}
allows us to rewrite the hypergeometric functions in the more
recognizable form
\begin{equation}
  \label{eq:Gqn}
  2\sin(\pi\mu)G(t,\Omega_{SP};t',\Omega_{SP})
  =e^{-i\pi h_+}R^{-h_+}H_+(P_\e)-e^{-i\pi h_-}R^{-h_-}H_-(P_\e)
\end{equation}
where $H_\pm$ are analytical continuations of the Green functions
$G_\pm$ we defined in \eqref{eq:Gpm} to the region $P>1$:
\begin{equation}
  H_\pm(x;x')=G_E(x;x')-e^{-i\pi h_\mp}G_E(x;x_A').
\end{equation}
They are explicitly given by
\begin{equation}
  H_{\pm}(P_\e)
  =\frac{\Gamma(\mp\mu)\Gamma(h_\pm)\sin(\pm\pi\mu)}{2^{1+2h_\pm}\pi^{5/2}}
  \pa{\frac{2}{1+P}}^{h_\pm}
  F\!\br{h_\pm,h_\pm-1;2\pa{h_\pm-1},\frac{2}{1+P_\e}}
\end{equation}
with the argument
\begin{equation}
  \label{eq:argument}
  \frac{2}{1+P_\e}\equiv\frac{4z}{(1+z)^2}
\end{equation}
while $R$ is some correction factor
\begin{equation}
  R\equiv z\pa{e^t-i\e}\pa{e^{-t'}-i\e}
  =e^{-(t-t')/2}\pa{e^t-i\e}^{1/2}\pa{e^{-t'}-i\e}^{1/2}.
\end{equation}
Note that when $\e\to0$, this correction $R\to1$ and $z\to
e^{-(t-t')}$.

Also, observe that \eqref{eq:argument} implies that
\begin{equation}
  \label{eq:Pe}
  P_\e=\frac{1}{2}\pa{z+\frac{1}{z}}
  =\frac{1}{2}\cu{
    \frac{e^{-(t-t')/2}}{\br{\pa{e^t-i\e}\pa{e^{-t'}-i\e}}^{1/2}}
    +\frac{\br{\pa{e^t-i\e}\pa{e^{-t'}-i\e}}^{1/2}}{e^{-(t-t')/2}}}.
\end{equation}
Away from the singularity at $t=t'$, we can set $\e$ to zero so that
\begin{equation}
  G(t,\Omega_{SP};t',\Omega_{SP})
  =\frac{e^{-i\pi h_+}H_+(P)-e^{-i\pi h_-}H_-(P)}{2\sin(\pi\mu)}
  =G_E(t,\Omega_{SP};t',\Omega_{SP}).
\end{equation}
The singularity structure for $G$ can be also analyzed from
\eqref{eq:Gqn}. The correction factor $R$ is regular near the
singularity, while the $G_{\pm}(P)$ have poles when $P$ approaches 1.
Expanding \eqref{eq:Pe} to first order in $\e$ yields
\begin{equation}
  P_\e=\cosh(t-t')
  -i\e\sinh(t-t')\pa{\frac{e^{-t}+e^{t'}}{2}}=P-i\e\hat{s}(x,x')
\end{equation}
with
\begin{equation}
  \hat{s}(x,x')\equiv\sinh(t-t')\pa{\frac{e^{-t}+e^{t'}}{2}}.
\end{equation}
The singularity structure therefore is the same as in definition
\eqref{eq:massiveGE} for the Euclidean Green function up to redefinition
of $\e$ by some positive function.

\end{document}